\documentclass[12pt]{article}
\usepackage{amssymb,amsmath}
 
\newcommand{\ket}[1]{\left|#1\right\rangle}
\newcommand{\bra}[1]{\left\langle#1\right|}

\newcommand{\numEq}[2]{\begin{equation}
   \label{eq:#1}
   #2\end{equation}}

\let\epsilon\varepsilon

\numberwithin{equation}{section}

\begin{document}

\title{\bf \mbox{}\vspace{-0.3in}\mbox{}\\
The $q$-deformed Jaynes-Cummings Model 
 and its $q$-Supercoherent States
 \thanks{This work was supported in part by the National Foundation of the Science 
 and Technology committee and alse by the Foundation of Science and 
Technoloy Committee of Hunan}}
\markright{The $q$-deformed Jaynes-Cummings Model and its $q$-Supercoherent
States}
\author{\normalsize San-Ru Hao\thanks{Email address: srhao@sparc2.hunnu.edu.cn} 
and  Lu-Ya Wang
\\ Calculating Physics Division, Department of Computer Teaching,\\
Human Normal University, Hunan 410081,People's Republic of China}
\date{}

\maketitle
\thispagestyle{empty}

\baselineskip 0.3in

\begin{abstract}\noindent\normalsize 
In this paper, we have proposed a $q$-deformed Jaynes-Cummings(JC) model 
and constructed the $q$-SuperCoherent States(q-SCSs) for the $q$-deformed JC 
model. We have also discussed the properties of the $q$-supercoherent states 
and given the completeness relation expression. The representation of the 
q-supercoherent states for the $q$-deformed JC model is studied as 
well. \\
\\
{\bf PACS number(s):} 03.65.Nk \\
{\bf Key Works:} $q$-deformed JC model, $q$-supercoherent states,$q$-SCSs 
representation.

\end{abstract}

\newpage

\baselineskip 0.3in
\indent

\section{\normalsize Introduction}\label{sec:1}

It is well known that the usual coherent states[1-4] of Lie(super)algebra 
have wide applications to various branches of physics. For example, the  
coherent property of the quantum state is very important to the quantum 
system which is used by the 
quantum computer[5-8]. Over past few years, quantum groups and theirs 
representations[9-10] have drawn considerable attention from mathematicians and 
physicists. People have focused much attention on the algebra of $q$-boson 
oscillators[11,12] for a few years. The q-boson operator algebra has been 
used as a tool for constructing the highest weight representation 
of $SU_q(2)$[11,12] and other quantum groups. 
The $q$-boson coherent states have been studied [13,14] and have 
shown interesting squeezing properties. Parthasarathy and Viswanathan [15] have 
proposed an algebra of $q$-fermion creation and annihilation operators, which allows 
(for $0<q<1$) any numbers of $q$-fermions to occupy a given states in contrast 
to the case of ordinary fermions. 
\par
It is natural to investigate the properties of the $q$-deformed quantum 
mechanics systems, which have the supersymmetric properties, and further 
to study the $q$-SuperCoherent States($q$-SCSs) of the $q$-deformed quantum 
supersymmetric systems by means of the $q$-boson and $q$-fermion operators 
algebra mentioned above.
\par
In our past work , we have studied many supersymmetric quantum system[16-20], 
but in this paper, we will propose a $q$-deformed scheme for the simplest 
version of Fermi-Bose interation --the Jaynes-Cummings(JC) model and consider 
the problem of constructing the $q$-SCSs for the $q$-deformed JC model. 
We will also discuss the properties of the q-SCSs as well.

\section{\normalsize $q$-deformed JC Model and $q$-supercoherent States}\label{sec:2}

\par
The JC model [21] is assumed to be the simplest version of matter-radiation . 
It describes a two-level atom (or a single ${\frac{1}{2}}$-spin)  coupled 
linearly with a single bosonic model. The model can be used in the study of 
the quantum computer, we will discuss this problem in our another work. 
\par
The Hamiltonian for JC model is 
\numEq{2.1}{ H_{JC}=2\omega_1b^+b+2\omega_2f^+f+\psi bf^+
        +b^+f\bar{\psi} \ ,
 }
where $b(b^+)$ and $f(f^+)$ are the bosonic and fermionic annihilation
(creation)operators, respectively, which satisfy the usual commutation and 
anticommutation relations
\numEq{2.2}{\begin{align}
            [b,b^+]=1, \hspace{0.5cm} \{f,f^+\}=1 , \\
            [b,f]=[b,f^+]=0, etc  . \nonumber
\end{align}
}
Here an atom is supposed to be in the eignstates which have the energies   
 $E=2\omega$ and $E=0$. The interaction constants $\psi$ and $\bar{\psi}$  
in equation (2.1) are the grassmann numbers. It is obvious that the Hamiltonian 
(2.1) is Hamitian and it describes the supersummetric quantum mechanics 
systems. 
\par
A problem of interest is the consideration of the $q$-deformed JC model, 
which describes the $q$-deformed supersymmetric quantum mechanics systems. 
Parthasarathy and Viswanathan [15] have proposed an algebra of $q$-fermions 
creation and annihilation operators. Doing as the same as Ref.[15], we 
propose that the $q$-deformed Hamiltonian $H_{JCq}$ of the JC model is given 
by the following expression
\numEq{2.3}{ 
    H_{JCq}=2\omega_1aa^++2\omega_2q^{\frac{M}{2}}f_q^+f_q+\psi_qaf_q^+q^{
     \frac{M}{4}}+q^{\frac{M}{4}}a^+f_q\bar{\psi}  ,
}
where $a$ and $a^+$ are the $q$-deformed bosonic operators satisfying 
the following commutation relations
\numEq{2.4}{
      aa^+-qa^+a=q^{-N},\hspace{0.5cm} [N,a]=-a,\hspace{0.5cm} [N,a^+]=a^+ ,
}
and where $f_q$ and $f_q^+$ are the $q$-deformed fermionic annihiltation 
and creation operators respectively, obeying the relations[15]
\numEq{2.5}{
    f_qf_q^++q^{\frac{1}{2}}f_q^+f_q=q^{-\frac{M}{2}} ,
    \hspace{0.5cm} [M,f_q^+]=f_q^+ ,\hspace{0.5cm} [M,f_q]=-f_q  .
}
As usual, we have 
\numEq{2.6}{
   [a,f_q]=[a,f_q^+]=0 , etc   .
} 
\par
In above equations,$N$ and $M$ are the bosonic and fermionic number 
operators respectively. And in equation (2.3), $\psi_q$ and $\bar{\psi_q}$ 
are taken as pseudo-Grassmann variables, which means that $\psi_q
\bar{\psi_q}+\bar{\psi_q}\psi_q=0$, and are not nilpotent, i.e. 
$(\psi_q)^n\not=0, (\bar{\psi_q})^n\not=0 $ for $n>1$. Further $\psi_q$  
and $\bar{\psi_q}$ are taken to anticommute with the $q$-deformed 
fermionic operators, but commute with the fermionic number operater $M$. 
\par
As it was shown in [15] that for $0<q<1$ any number of $q$-fermions can occupy 
a gevien state in contrast to the case of ordinary fermions. For this we 
suppose that the $q$-fermionic operators in (2.3), which satisfy the relations 
(2.5) and (2.6), are not nilpotent for $0<q<1$, but the nilpotent relations   
$f_{q=1}^2=0$ and $(f_{q=1}^+)^2=0$ are still satisfied in the weak sense 
when $q=1$. 
\par
For simplicity, we can define in eq.(2.3) the following transformations[20,21]:
\numEq{2.7}{
      f_q=q^{-\frac{M}{4}}F ,\hspace{0.5cm} f_q^+=F^+q^{-\frac{M}{4}}   ,
}
then, the $q$-deformed JC model(2.3) can be expressed as follows:
\numEq{2.8}{
  H_{JCq}=2\omega_1a^+a+2\omega_2q^{\frac{1}{2}}F^+F+\psi_qaF^++a^+F\bar{\psi_q}  .
}
The basic anticommutation relations become
\numEq{2.9}{
  FF^++qF^+F=1,\hspace{0.5cm} [M,F^+]=F^+, \hspace{0.5cm} [M,F]=-F .
}  
\par
In order to construct the $q$-supercoherent states for the $q$-JC model, we 
first construct a super-Fock space based on the vaccum state
\numEq{2.10}{
  \ket{0}=\ket{0}_B\otimes\ket{0}_F  , 
  }
where $\ket{0}_B$ and $\ket{0}_F$ are the vacuum states of $q$-bosonic 
and $q$-fermionic operators respectively, i.e. 
\numEq{2.11}{
  a\ket{0}_B=0,\hspace{0.5cm} F\ket{0}_F=0  .
}
Iterating (2.9), one arrive at the formula
\numEq{2.12}{
  F(F^+)^n-(-q)^n(F^+)^nF=[n]_F(F^+)^{n-1}  ,
}
where $[n]_F$ is the fermionic $q$-number defined by the expression
\numEq{2.13}{
  [n]_F={\frac{(1-(-q)^n)}{(1-(-q))}} .
  }
Making use of the same way, but interating (2.4), one can obtain the 
following expression as well
\numEq{2.14}{
  a(a^+)^n-q^n(a^+)^na=[n]_B(a^+)^{n-1}q^{-N}  ,
}
where $[n]_B$ is the bosonic $q$-number defined as follows
\numEq{2.15}{
  [n]_B={\frac{(q^n-q^{-n})}{(q-q^{-1})}} . 
}  
\par
By making use of the vacuum state (2.10) and by considering the equation 
(2.11), the normalized $n$ $q$-bosons and $m$ $q$-fermions states can be 
constructed , which is expressed as the follows:
\numEq{2.16}{
   \ket{n,m}_q={\frac{(a^+)^n(F^+)^m}{\sqrt{[n]_B![m]_F!}}}\ket{0}  ,
}
where
\numEq{2.17}{\begin{align}
  [n]_B!=[n]_B[n-1]_B\cdots [1]_B  , \\
  [m]_F!=[m]_f[m-1]_F\cdots [1]_F  .  \nonumber
\end{align}
}
From (2.16), and by making use of(2.12) and (2.14), the action of the 
$q$-boson operators $a(a^+)$ and the $q$-fermion operators $F(F^+)$ on 
the base vectors in the super-Fock space can be given by
\numEq{2.18}{\begin{align}
    & a\ket{n,m}_q=\sqrt{[n]_B}\ket{n-1,m}_q \nonumber ,\\
    & a^+\ket{n,m}_q=\sqrt{[n+1]_B}\ket{n+1,m}_q \nonumber,\\
    & F\ket{n,m}_q=\sqrt{[m]_F}\ket{n,m-1}_q ,           \\
    & F^+\ket{n,m}_q=\sqrt{[m+1]_F}\ket{n,m+1}_q \nonumber  .
\end{align}
}
From equations(2.12) and (2.14), and making use of an induction method, one 
can get the orthogonality relations
\numEq{2.19}{
  _q\bra{n,m}n',m'>_q=\delta_{nn'}\delta_{mm'} .  
}
\par
In order to construct the $q$-supercoherent states of the $q$-JC model, 
we introduce two $q$-exponential functions for the $q$-bosonic operators 
and the $q$-fermionic operators respectively, which are defined as 
\numEq{2.20}{\begin{align}
  & \exp_{Bq}(zq^+)=\sum\limits_{n=0}^{\infty}{\frac{z^n(a^+)^n}{[n]_B!}}  ,\\
  & \exp_{Fq}(\psi_qF^+)=\sum\limits_{m=0}^{\infty}{\frac{(\psi_qF^+)^m}{[m]_F!}} \nonumber .
\end{align}
}
These $q$-exponential functions are the $q$-analog of the usual classical 
ones. In equation (2.20) , $z$ is taken as the complex number and the 
$\psi_q$, which is introduced by equation (2.8), is a pseudo-Grassmann 
variable and is not nilpotent, but $\psi_q \longrightarrow\psi$ if $q
\longrightarrow 1$. 
\par
Let us now construct the $q$-SCSs. The $q$-SCSs of the $q$-JC model can 
be defined as
\numEq{2.21}{
  \ket{z,\psi_q}=N(\bar{z}z,\psi_q\bar{\psi}_q)\exp_{Bq}(za^+)
     \exp_{Fq}(-\psi_qF^+)\ket{0} ,
}
where $N(\bar{z}z,\psi_q\bar{\psi}_q)$ is the normalization factor, and the 
$q$-exponential functions are defined by (2.20). Since $z$ is a complex 
variable and $\psi_q$ has the properties of the Grassmann varibles, 
then we call $\ket{z,\psi_q}$ as the $q$-supercoherent states.

\section{\normalsize  Properties of $q$-supercoherent states}

\par
There are essentially two properties that all coherent states have to share 
in common. The first property is the continuity , i.e. the coherent state is 
a strongly continuous function of its label variables; The second property 
that all sets of coherent states share in common is the completeness (or 
called as resolution of unity), i.e. there exists a positive measure $d\sigma$ 
on the space in which the coherent state is defined so that the unit operator  
$I$ admits the resolution of unity,
\numEq{3.1}{
  I=\int d\sigma \ket{x}\bra{x}   ,
}
where $\ket{x}$ is a coherent state.
\par
For the q-SCSs $\ket{z,\psi_q}$, it must has the two properties mentioned 
above. Now we discuss those properties detailed.
\par
By means of Eqs.(2.20), the q-SCSs (2.21) can be rewritten as
\numEq{3.2}{\begin{align}
   \ket{z,\psi_q} & =N(\bar{z}z,\bar{\psi}_q\psi_q)\sum\limits_{n,m=0}^{\infty}
                     {\frac{z^n(a^+)^n(-\psi_qF^+)^m}{[n]_B![m]_F!}}\ket{0}, \nonumber \\
                  & =N(\bar{z}z,\bar{\psi}_q\psi_q)\sum\limits_{n,m=0}^{\infty}
                     {\frac{(-1)^{<m/2>+m}z^n(\psi_q)^m}{\sqrt{[n]_B![m]_F!}}}
                     \ket{n,m}_q ,
\end{align}
}
where $<m/2>$ stands for the integer part of $m/2$. We have used equation 
(2.16) in the second equality of equation (3.2).
\par
Let us consider the normalized q-SCSs. Thus we have 
\numEq{3.3}{
   \bra{z,\psi_q}z,\psi_q>=1  .
}   
\par
Substituting (3.2) into (3.3) and using (2.19), one can obtain the normalized 
form of the q-SCSs
\numEq{3.4}{
    \bra{z,\psi_q}z,\psi_q>=N^2(\bar{z}z,\bar{\psi}_q\psi_q)
              \exp_{Bq}(\bar{z}z)\exp_{Fq}(\bar{\psi}_q\psi_q)=1  .
}
Then the normalization constant (or factor) is given by the expression
\numEq{3.5}{
     N(\bar{z}z,\bar{\psi}_q\psi_q)=(\exp_{Bq}(\bar{z}z)\exp_{Fq}
            (\bar{\psi}_q\psi_q))^{-1/2}   .
}
\par
In view of (3.2) and the inner product (2.19), it follows that the 
$q$-supercoherent states overlap is given by
\numEq{3.6}{
    \bra{z,\psi_q}z',\psi_q'>=N(\bar{z}z,\bar{\psi}_q\psi_q)
          N(\bar{z}'z',\bar{\psi}_q'\psi_q')\exp_{Bq}(\bar{z}z')
          \exp_Fq(\bar{\psi}_q\psi_q')   .
}
It means that the $q$-SCSs of the $q$-deformed JC model are not orthogonal 
to each other and that they are linearly dependent. It is obvious that the 
$q$-SCSs are the strongly contiuous function of their label variables $z$   
and $\psi_q$. As is mentioned above, the core of coherent states is their 
completeness(in another words: resolution of unity).
\par
As is well known, the resolution of unity in the super-Fock space spanned by 
the base vectors defined by (2.16) can be written as
\numEq{3.7}{
   I=\sum\limits_{n,m=0}^{\infty}\ket{n,m}{_q}{_q} \bra{n,m}   .
}
By means of this equation, in the present case, it can be shown that the 
$q$-SCSs form a complete set. And their completeness relation takes the form
 \numEq{3.8}{
   \int\int d_q^2zd_{Fq}^2\psi_q\sigma(\bar{z}z,\bar{\psi}_q\psi_q)   
   \ket{z,\psi_q}\bra{z,\psi_q}
   =\sum\limits_{n,m=0}^{\infty}\ket{n,m}{_q}{_q} \bra{n,m}=I ,
 }
where
\numEq{3.9}{\begin{align}
  & d_q^2z={\frac{1}{[2]_B}}d_qr^2d\varphi,\hspace{0.5cm}  z=re^{i\varphi} , 
          \hspace{0.5cm} d_{Fq}^2\psi_q=d_{Fq}(\bar{\zeta}\zeta)d\theta , \\
  & \psi_q=\zeta e^{i\theta} ,\hspace{0.5cm} \bar{\psi}_q=\bar{\zeta}e^{-i\theta} ,
  \hspace{0.5cm} \bar{\zeta}\zeta+\zeta\bar{\zeta}=0 \nonumber .
\end{align}
}
Note that $\psi_q$ is the pseuo-Grassmann variable and the integral over   
$d_qr^2$ and $d_{Fq}(\bar{\zeta}\zeta)$ are the $q$-integration. while the 
integral over $d\varphi$ and $d\theta$ are the usual integration. The weight 
$q$-super-function $\sigma(\bar{z}z,\bar{\psi}_q\psi_q)$ in equation (3.8) 
is given by
\numEq{3.10}{
  \sigma(\bar{z}z,\bar{\psi}_q\psi_q)={\frac{[2]_B\exp_{Bq}(-\bar{z}\bar{z})
        \exp_{Fq}(\bar{\psi}_q\bar{\psi}_q)}
        {(2\pi)^2\exp_{Bq}(\bar{z}z)\exp_{Fq}(\bar{\psi}_q\psi_q)}} .
}
\par
We now prove the completeness relation (3.8). Substituting (3.2),(3.5) and 
(3.10) into (3.8), one can obtain
\numEq{3.11}{\begin{align}
 & l.h.s \hspace{0.3cm} of (3.8)  \nonumber \\
 &=\sum\limits_{n,m=0}^{\infty}\sum\limits_{n',m'=0}^{\infty}
   {\frac{[2]_B}{(2\pi)^2}}\int\int d_q^2zd_{Fq}^2\psi_q\exp_{Bq}(-\bar{z}z)
   \exp_{Fq}(\bar{\psi}_q\psi_q)  \nonumber \\
 &\hspace{0.3cm}\cdot{\frac{(-1)^{<m/2>+<m'/2>+m+m'}}{\sqrt{[n]_B![m]_F!
   [n']_B![m']_F!}}}(\psi_q)^m(\bar{\psi}_q)^{m'}\ket{n,m}{_q}{_q}\bra{n',m'} \nonumber \\
 &=\sum\limits_{n,m=0}^{\infty}\sum\limits_{n',m'=0}^{\infty}
    {\frac{[2]_B}{(2\pi)^2}}\int\int ([2]_B)^{-1}d_qr^2d\varphi d_{Fq}(\bar{\zeta}\zeta)
    d\theta\exp_{Bq}(-r^2)\exp_{Fq}(\bar{\zeta}\zeta) \nonumber \\
 &\hspace{0.3cm}\cdot {\frac{(-1)^{<m/2>+<m'/2>+m+m'}}{\sqrt{[n]_B![m]_F!
   [n']_B![m']_F!}}}e^{i(n-n')\varphi}e^{i(m-m')\theta}r^{m+m'}\zeta^m
    \bar{\zeta}^{m'}\ket{n,m}{_q}{_q}\bra{n,m} \nonumber  \\
 &=\sum\limits_{n,m=0}^{\infty}\int d_{Fq}(\bar{\zeta}\zeta)
     {\frac{(\bar{\zeta}\zeta)^m}{[m]_F!}}\exp_{Fq}(-\bar{\zeta}\zeta)
     \ket{n,m}{_q}{_q}\bra{n,m}   .
\end{align}
}
where we have used the $\varphi$ and $\theta$ integration
\numEq{3.12}{
   {\frac{1}{2\pi}}\int\limits_0^{2\pi}\exp(i(n-m)\varphi)d\varphi=
     {\frac{1}{2\pi}}\int\limits_0^{2\pi}\exp(i(n-m)\theta)d\theta 
     =\delta_{nm}  ,
}
and the $q$-Euler formula for the function $\Gamma(x)$[12]
\numEq{3.13}{
  \int\limits_0^{\xi}exp_{Bq}(-x)d_qx=[n]_B!  ,
}
for $\xi$ being a large number.
\par
For quantum groups, we define here the $Fq$-derivative for the 
pseudo-Grassmann variables as
\numEq{3.14}{
  {\frac{df(x)}{d_{Fq}x}}={\frac{f(x)-f(-qx)}{x-(-qx)}}  ,
}
hence, one can obtain the $Fq$-Euler formula for the function $\Gamma(x)$:
\numEq{3.15}{
  \int\limits_0^{\xi}x^n\exp_{Fq}(-x)d_{Fq}x=[n]_F!  ,
}
where $\xi$ is a large number.
\par
By making use of (3.15), it follows from (3.11) that 
\numEq{3.16}{
  l.h.s\hspace{0.3cm} of (3.8)=\sum\limits_{n,m=0}^{\infty}\ket{n,m}{_q}
    {_q}\bra{n,m}=I  .
}
Thus, we have proved that the q-SCSs $\ket{z,\psi_q}$  defined by (2.21) 
are completeness and equation (3.8) holds.
\par
With the aid of the completeness relation of the q-SCSs , one can expand an 
arbitrary vector $\ket{f}$ as 
\numEq{3.17}{
   \ket{f}=\int\int d_q^2zd_{Fq}^2\psi_q\sigma(\bar{z}z,\bar{\psi}_q\psi_q)
      \ket{z,\psi_q}\bra{z,\psi_q}f>  .
}
Setting $\ket{f}=\ket{z',\psi_q'}$, an arbitrary $q$-SCSs of the $q$-JC model, 
then we have
\numEq{3.18}{
  \ket{z',\psi_q'}=\int\int d_q^2zd_{Fq}^2\psi_q\sigma(\bar{z}z,\bar{\psi}_q\psi_q)
    \ket{z,\psi_q}\bra{z,\psi_q}z',\psi_q'>  .
}
This means that the system of q-supercoherent states is over-complete.

\section{\normalsize $q$-SCSs representation of the $q$-deformed JC model}

\par
In this section, we study the $q$-SCSs representation of the $q$-deformed 
JC model $H_{JCq}$ defined by (2.8). The trace of $H_{JCq}$ is discussed as 
well. These quantities have widely been used in constructing the coherent 
state path-integral representation of models and in studying the decoherence 
of the quantum system based by a quantum computer.
\par
From equations (2.18) and considering (3.2), we arrive at the formulas
\numEq{4.1}{
  a\ket{z,\psi_q}=z\ket{z,\psi_q} , \hspace{0.8cm} F\ket{z,\psi_q}=\psi_q\ket{z,\psi_q}  .
}
So we further have
\numEq{4.2}{
  \bra{z,\psi_q}=\bra{z,\psi_q}\bar{z} , \hspace{0.8cm} \bra{z,\psi_q}=
    \bra{z,\psi_q}\bar{\psi}_q  .
}
If one hope to use the standard path-integral method in studying JC model 
and related problems, the coherent states representation of the Hamiltonian 
must be used. Therefore we will study the $q$-SCSs representation of the 
$q$-deformed JC model here. 
From (2.8) and by making use of (4.1) and (4.2), it follows that the $q$-SCSs 
representation of $H_{JCq}$ can be given by 
\numEq{4.3}{
  H_{JCq}(z,\psi_q)\equiv\bra{z,\psi_q}H_{JCq}\ket{z,\psi_q}  
        =2\omega_1\bar{z}z+2\omega_2q^{\frac{1}{2}}\bar{\psi}_q\psi_q
                   +(\bar{z}+z)\psi_q \bar{\psi}_q .
}
It is obvious that (4.3) is the diagnonal matrix element of the $q$-deformed 
JC Hamiltonian operator $H_{JCq}$. The off-diagnonall matrix elements of 
operator $H_{JCq}$, which is very interested to some peoples who are studying 
the decoherence of the quantum computer system[8], can be expressed by 
\numEq{4.4}{\begin{align}
   & H_{JCq}(z,\psi_q;z',\psi_q') \equiv \bra{z,\psi_q}H_{JCq}\ket{z',\psi_q'} \nonumber \\
   & \hspace{0.5cm}=N(\bar{z}z,\bar{\psi}_q\psi_q)N(\bar{z}'z',\bar{\psi}_q'\psi_q')
     \exp_{Bq}(\bar{z}z')\exp_{Fq}(\bar{\psi}_q'\psi_q)  \nonumber \\
   & \hspace{0.8cm}\cdot(2\omega_1\bar{z}z'+2\omega_2q^{\frac{1}{2}}\bar{\psi}_q\psi_q'
     +\psi_q\bar{\psi}_qz'+\bar{z}\psi_q'\bar{\psi}_q)  .  
\end{align}
}
Apart from the normalization factors, this defines an entire function of 
variables $z$,$z'$,$\psi_q$ and $\bar{\psi}_q$. It is straightforward to 
see that such a function is uniquely determined by its diagonal 
representation (4.3).
\par
Making use of the completeness relation (3.8), we observe that the trace 
of $H_{JCq}$ may be represented by 
\numEq{4.5}{
  Tr(H_{JCq})=\int\int d_q^2zd_{Fq}^2\psi_q\sigma(\bar{z}z,\bar{\psi}_q\psi_q)
      H_{JCq}(z,\psi_q)  ,
}
which is a useful expression for constructing the $q$-SCSs path-integral 
representation of this model and for discussing the decoherence of the 
quantum computer system, and we will use this expression in our another work.

\section{\normalsize Conculding remark}

\par
We have proposed a $q$-deformed scheme for the usual JC model and 
constructed the $q$-supercoherent states for this model. We have discussed 
the properties of the $q$-supercoherent states as well. The $q$-supercoherent 
states have some advantages such as they contain the usual complex variables 
and the Grassmann variables, so that one can easily apply them to the 
supersymmetric quanmtu systems. 
\par
It is possible to generalize this kind of 
$q$-supercoherent states to other supersymmetric quantum systems, e.g. the 
$q$-superoscillators in supersymmetric quantum systems. It is also 
interesting to discuss the decoherence of the quantum computer systems by 
using the representation of the $q$-SCSs of the $q$-deformed JC model. 
These two works are done by us , the reader who are interested in those can 
read our works[8,16-20].

\section{\normalsize Acknowledgement}

\par
This research was supported by the National Foundation of the Science and 
Technology Committee, and also by the Foundation of the Science and 
Technology Committee of Hunan.

\end{document}